\documentclass[aps,prl,showpacs,twocolumn]{revtex4}
\usepackage{amssymb}
\usepackage{amsmath}
\usepackage{graphicx}
\usepackage{color}

\begin{document}

\title{Comparative study of magnetic quantum oscillations in Hall and transverse
magnetoresistance}
\author{A.A.~Sinchenko$^{1}$, P.D.~Grigoriev$^{2,3,4}$, A.V.~Frolov$^4$, A.P.~Orlov$^4$, V.N.~Zverev$^5$, A.~Hadj-Azzem$^{6}$, E.~Pachoud$^{6}$ and P.~Monceau$^{6}$}

\address{$^{1}$Laboratoire de Physique des Solides, Universit´e Paris-Saclay, CNRS, 91405 Orsay Cedex, France}

\address{$^{2}$L. D. Landau Institute for Theoretical Physics, 142432,
	Chernogolovka, Russia}

\address{$^{3}$National University of Science and Technology "MISiS", 119049
	Moscow, Russia}

\address{$^{4}$Kotelnikov Institute of Radioengineering and Electronics of RAS, 125009 Moscow, Russia}

\address{$^{5}$Institute of Solid State Physics, Chernogolovka, Moscow region 142432, Russia}

\address{$^{6}$Univ. Grenoble Alpes, Inst. Neel, F-38042 Grenoble, France,
	CNRS, Inst. Neel, F-38042 Grenoble, France}

\begin{abstract}
 Magnetic quantum oscillations (MQO) of Hall coefficient are measured in rare-earth 
 tritelluride TmTe$_{3}$ and shown to be much stronger and persist to higher 
 temperature than the Shubnikov oscillations. It is general for MQO in strongly anisotropic metals, and 
the combined measurements of Hall and diagonal magnetoresistance provide 
useful informations about the electronic structure. The ratio of their MQO amplitudes 
depends linearly on magnetic field, and its slope gives a simple and accurate measurement tool
of the electron mean free time and its temperature dependence. 
\end{abstract}

\date{\today}

\maketitle

The Landau quantization of electron spectrum in magnetic fields leads to the
magnetic quantum oscillations (MQO) in metals \cite{Shoenberg,Abrik,Ziman}. 
Usually, the MQO are observed in magnetoresistance, called the Shubnikov-de
Haas effect (ShdH), and in magnetization, called the de Haas-van Alphen
effect (dHvA). These quantities are measured as a function of the inverse
magnetic field and display a periodic behavior. The period is given by the
extremal cross section of the Fermi surface (FS) encircled by conducting
electrons in a semiclassical picture. The amplitude of the MQO is given by
the well known Lifshitz-Kosevich (LK) formula \cite{LK}. This formula gives
the relation between the MQO frequency and FS, and describes the MQO damping
by thermal and disorder broadening. Fitting the experimental temperature
dependence of MQO amplitude by the LK formula gives the effective electron
mass $m^{\ast}$, while the field dependence of MQO amplitude gives the
Landau-level (LL) broadening \cite{Shoenberg}. The MQO measurements provide a
powerful tool to study the electronic properties of various
quasi-two-dimensional (Q2D) layered metallic compounds which are the subject
of intense studies now: organic metals \cite{OMRev,MarkReview2004}, cuprate
and iron-based high-temperature superconductors \cite%
{ProustNature2007,DVignolle2008,HelmNd2009,HelmNd2010,SebastianPRL2012,Graf2012,BaFeAs2011,Coldea2013},
heterostructures \cite{Kuraguchi2003,Stormer2007}, graphite intercalation compounds \cite%
{Csanyi2003}, various van-der-Waal crystals \cite{Fallahazad2016}, topological semimetals \cite{Busch2018,Jiang2021}, etc.

\begin{figure}[bt]
\includegraphics[width=8.5cm]{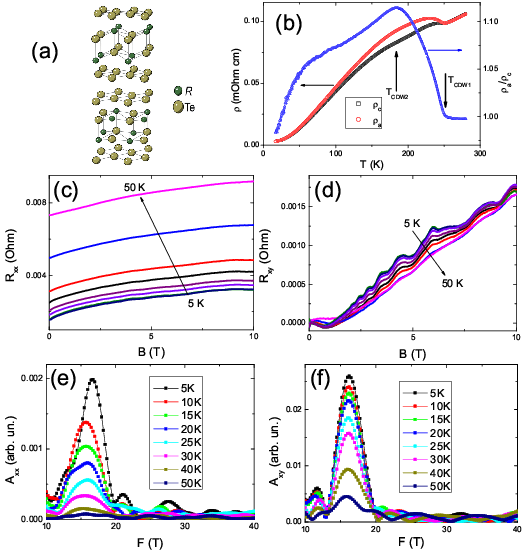}  
\caption{(color online)(a) Crystal structure of RTe$_3$ compounds. (b)
Temperature dependence of resistivity of TmTe$_3$ along the $a$ and the $c$
-axis directions and conductivity anisotropy, $\protect\rho_a/\protect\rho_c$,
 in the $a-c$ plane. (c) Magnetoresistance, $R_{xx}$, and (d) Hall
resistance, $R_{xy}$, in TmTe$_3$ as a function of magnetic field, $B$ applied pependicular to the ($a-c$) plane, at
various temperatures in the range 5-50 K. Panels (e) and (f) show the corresponding
FFT.}
\label{F1}
\end{figure}

Usually, only the diagonal component of magnetoresistance tensor is used to
measure the MQO and to study the electronic structure, although the MQO of
non-diagonal Hall component are also clearly observed and may even be
stronger. For example, the MQO in the hole-doped high-Tc
cuprate superconductors were first discovered measuring the Hall resistance 
\cite{ProustNature2007}. The high-temperature quantum oscillations of the Hall 
resistance were also measured in topological semimetals, such as bulk Bi$_2$Se$_3$ 
\cite{Busch2018}. It seems interesting to compare the MQO of diagonal
and Hall magnetoresistance components and to analyze if that gives
additional useful informations about the electronic structure. For this
purpose we choose a Q2D compound TmTe$_{3}$. Compounds of family $R$Te$%
_{3}$ ($R=$Y, La, Ce, Nd, Sm, Gd, Tb, Ho, Dy, Er, Tm) have weak orthorhombic
structure (space group $Cmcm$) in the normal state (see Fig. \ref{F1}(a)). These
systems exhibit a $c$-axis incommensurate charge-density wave (CDW) at high
temperature through the whole $R$ series that was recently a subject of
intense studies \cite{DiMasi95,Brouet08,Ru08,SinchPRB12,Anis14,SSC14}. For
the heaviest rare-earth elements, a second $a$-axis CDW occurs at low
temperature. MQO in RTe$_{3}$ compounds have been studied in works \cite%
{Borzi08,Sinch2016,Lei20,Walmsley20}. It was shown in \cite{Walmsley20} that in
RTe$_{3}$ compounds with the double charge density wave state several small
pockets with a very small effective mass and with the largest occupying
around 0.5\% of the Brillouin zone remain. TmTe$_{3}$ is a member of RTe$%
_{3} $ family with the heaviest rare-earth element and demonstrates the
lowest transition temperature $T_{CDW1}=250$ K of first high-T CDW and the
highest transition temperature $T_{CDW2}=190$ K of the second low-T CDW \cite%
{Ru08}. Hence, TmTe$_{3}$ is most convenient for the comparative study of
MQO in Hall and diagonal magnetoresistance in a wide temperature range.

Single crystals of TmTe$_3$ were grown by a self-flux technique under
purified argon atmosphere as described previously \cite{SinchPRB12}. Thin
single-crystal samples with a rectangular shape and with a thickness
typically 1-2 $\mu $m were prepared by micromechanical exfoliation of
relatively thick crystals glued on a sapphire substrate. RTe$_3$ compounds
are quite sensitive to air, so the crystals should be stored in an oxygen
and moisture free environment and all manipulation with the crystals in air
should be done during minimal time. Because of this feature the electrical
contacts were prepared by cold soldering of In. The magnetic field was
applied parallel to the $b$ axis, and in-plane magnetoresistance and the
Hall resistance were recorded using the van der Pauw method \cite{Pauw61},
sweeping the field between $+10$ and $-10$ T. Measurements were performed at
fixed temperature in the temperature range 4.2-100 K. Magnetic field
dependencies of resistance and Hall resistance were determined as $\dfrac{%
(V(+B)\pm V(-B))}{2I}$ taking ($+$) for magnetoresistance and ($-$) for Hall
resistance correspondingly. Conductivity measurements were performed using
the Montgomery technique \cite{Montgomery71,Logan71}.

Figure \ref{F1}(b) shows the temperature dependence of resistivity of TmTe$%
_{3}$ measured along the in-plane $c$ and $a$ axes together with the
anisotropy ratio $\rho _{a}/\rho _{c}$ in the conducting $ac$ plane using
the Montgomery method. Above the Peierls transition temperature $T_{CDW1}=270
$ K the studied compound is practically isotropic in the $ac$ plane and $%
\rho _{a}/\rho _{c}\approx 1$. Below $T_{CDW1}$ the ratio $\rho_{a}/\rho _{c}
$ strongly increases in agreement with Ref. [\onlinecite{Anis14}]. Below the
second CDW transition temperature the resistivity anisotropy decreases, and
at $T<80$ K it becomes less than 5\%. In this temperature range the compound
can be considered as nearly isotropic in $(ac)$ plane.

Figs. \ref{F1}c and \ref{F1}d show the diagonal $R_{xx}$ and Hall $R_{xy}$
transverse magnetoresistance components in TmTe$_3$ as a function of
magnetic field $B$ at various temperatures $T$ in the range 5-50 K. The MQO
of $R_{xy}$ are much more pronounced than of those of $R_{xx}$. Panels (e) and (f)
demonstrate corresponding Fourier transforms (FFT) in the window 3-9 T for
MQO of resistivity components $\rho_{xx}$ and $\rho_{xy}$. The MQO with
frequency $F=15$ T clearly manifest in both the diagonal and Hall
magnetoresistance. However, the MQO of Hall resistivity are much stronger
and observable till considerably higher temperature. This difference between
the MQO of Hall and diagonal magnetoresistance components has not been
previously pointed out.

The temperature dependence of MQO amplitude $A\left( T,B\right) $ is used to
extract the effective electron mass $m^{\ast }$, and its field dependence to
extract the Dingle temperature $T_{D}=\hbar /2\pi k_{B}\tau $, related
to the electron mean free time $\tau $, where  $k_{B}=1.38\cdot 10^{-16}$
erg/K is the Boltzmann's constant. In 2D metals the amplitude of MQO is
described by modified Lifshitz-Kosevitch formula \cite{Shoenberg84}: 
\begin{equation}
A\left( T,B\right) \propto R_{T}\left( T,B\right) R_{D}\left( B\right) ,
\label{LK}
\end{equation}%
where the temperature damping factor%
\begin{equation}
R_{T}=R_{T}\left( T,B\right) =\frac{\lambda }{\sinh (\lambda )},~\lambda
\equiv \frac{2\pi k_{B}T}{\hbar \omega _{c}},  \label{RT}
\end{equation}%
$\omega _{c}=eB/m^{\ast }c$ is the cyclotron frequency, 
$e$ is the electron charge, $c$ is the light velocity. The damping of MQO
by disorder is described by the usual Dingle factor%
\begin{equation}
R_{D}=\exp \left( -\frac{2\pi ^{2}k_{B}T_{D}}{\hbar \omega _{c}}\right)
=\exp \left( -\frac{\pi }{\omega _{c}\tau }\right) .  \label{RD}
\end{equation}

\begin{figure}[bt]
\includegraphics[width=8.5cm]{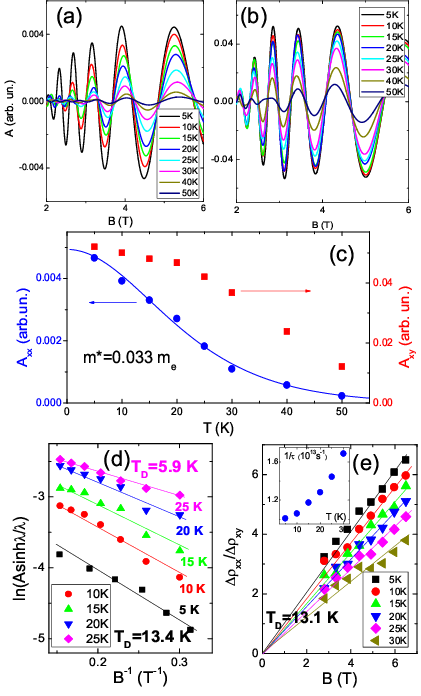}
\caption{(color online) Temperature evolution of the
MQO of magnetoresistance (a) and of Hall resistance (b) in TmTe$%
_3 $ for $F=15$ T. (c) The temperature dependence of the amplitude of Shubnikov oscillations
for $F=15$ T (blue symbols) and the corresponding Lifshitz-Kosevitch fits (blue
solid lines). Red squares indicate the temperature dependence of MQO amplitude of
the Hall resistance. (d) The Dingle plots, $\ln (A\sinh (\protect\lambda )/\protect\lambda %
)(B^{-1})$, for  the MQO of diagonal magnetoresistance $\protect%
\rho_{xx}$ at the same temperatures. (e) The magnetic-field dependence of the ratio of MQO
amplitudes, $A_{xx}/A_{xy}$, at various temperatures. Inset shows the
corresponding temperature dependence of the scattering rate, $1/\protect\tau (T)$.}
\label{F2}
\end{figure}

The magnetic oscillations of diagonal and Hall
magnetoresistance at different temperatures are shown in Fig. \ref{F2} (a)
and (b) correspondingly. Fig. \ref{F2}c demonstrates the temperature
evolution of MQO amplitudes. The MQO amplitude $A_{xx}$ of the diagonal
magnetoresistance (blue symbols) is well fitted by the Eq. (\ref{LK}) (blue
solid lines) with the best-fit value $m_{\alpha}^*=0.033m_e$. The low ShdH
frequency and very small effective mass indicate the existence of small FS
pockets with very light carriers in these compounds at low $T$ in agreement
with Ref. \cite{Walmsley20}. At the same time, the temperature dependence of
the oscillation amplitude $A_{xy}$ of Hall resistance, indicated by red
squares in Fig. \ref{F2}c, cannot be described by the same formula because $%
A_{xy}$ decreases much slower than $A_{xx}$ as temperature grows and the MQO
of $\rho_{xy}$ are observable up to much higher temperatures.

The Dingle temperature and the scattering time are, usually, extracted from
the so-called Dingle plot that is the logarithm of MQO amplitude divided by
a thermal damping factor, $\lambda/\sinh (\lambda)$, plotted as a function
of inverse magnetic field, $1/B$. The corresponding Dingle plots 
for the MQO of $\rho_{xx}$ are shown in Fig. \ref{F2} (d) at various
temperatures. We see that these plots and their slope change strongly as
temperature increases. The Dingle temperature extracted from the slope of
these curves at $T=5$ K is equal to $T_D\approx 13.4$ K, while at $T=25$ K
it decreases to $T_D\approx 5.9$ K. Of course, this strong decrease of $T_D
(T)$ is not physical and appears from the incorrect use of Eq. (\ref{LK})
beyond its applicability region. As we argue below, one may use the value $%
T_D$ extracted only at low temperature. The corresponding scattering times
extracted from the Dingle plot at low $T=5$ K is $\tau_\alpha =(0.90\pm
0.07)10^{-13}$ seconds.

The MQO of magnetoresistance in two-dimensional (2D) electron systems for
low/intermediate magnetic fields were theoretically studied in Ref. \cite%
{Isihara86}. According to this work, the MQO should be observable both in
diagonal and Hall magnetoresistance components, and for one-band 2D metals
they are given by simple formulas: 
\begin{equation}
\rho _{xx}=\frac{1}{\sigma _{0}}\left( 1+2\dfrac{\Delta g(T)}{g_{0}}\right) ,
\label{xx}
\end{equation}%
\begin{equation}
\rho _{xy}=\frac{\omega _{c}\tau }{\sigma _{0}}\left( 1-\dfrac{1}{%
(\omega _{c}\tau )^{2}}\dfrac{\Delta g(T)}{g_{0}}\right) ,  \label{xy}
\end{equation}%
where in a weak magnetic field, when high harmonics of MQO are small,%
\begin{equation}
\dfrac{\Delta g(T)}{g_{0}}=-2\cos \left( \frac{2\pi \varepsilon _{F}}{\hbar
\omega _{c}}\right) R_{D}R_{T}  \label{dg}
\end{equation}%
is the oscillatory part of the density of states (DoS), multiplied by the
temperature damping factor $R_{T}$, $\varepsilon _{F}$ is the Fermi energy,
and the damping factors $R_{D}$ and $R_{T}$ are given by Eqs. (\ref{RD}) and
(\ref{RT}). 
Eqs. (\ref{xx})-(\ref{dg}) were recently generalized \cite{Mogilyuk2018,GrigorievHallTheory2024} 
to layered quasi-2D metals (see Eqs. (57)-(60) of Ref. \cite{GrigorievHallTheory2024}) 
and shown to be valid in the main (first) order in the Dingle 
factor $R_D$ even at finite interlayer electron transfer integral $t_z$ if the oscillating DoS 
in Eq. (\ref{dg}) is multiplied by the additional factor $J_0 (4\pi t_z /(\hbar
\omega _{c}))$ typical to quasi-2D metals, where $J_0(x)$ is the Bessel function of zeroth order. 

Now the fact that the observed MQO in Hall resistance $\rho _{xy}$ are 
stronger and observable up to much higher
temperatures than the MQO of $\rho _{xx}$ is not surprising because it
directly follows from Eqs. (\ref{xx}) and (\ref{xy}). 
Indeed, in contrast to MQO of $\rho _{xx}$ the amplitude $A_{xy}$ of MQO in $\rho
_{xy}$ is inversely proportional to $\tau_0$, which should decrease as 
the temperature grows because the electron levels become broadened 
not only by static crystal disorder but also by thermal excitations 
due to the  electron-phonon (e-ph) and electron-electron (e-e) interaction. 

We now emphasize another interesting point: Eqs. (\ref{xx}) and (\ref{xy}) 
predict a very simple formula for the ratio of MQO amplitudes, 
\begin{equation}
\Delta \rho _{xx}/\Delta \rho _{xy}=2\omega _{c}\tau = 2e B \tau /(m^{\ast }c).
 \label{ratio}
\end{equation}%
 Hence, plotting 
the ratio $\Delta \rho _{xx}/\Delta \rho _{xy}$ as a function of magnetic 
field $B$ one obtains a linear dependence with a slope equal to 
$2e\tau /(m^{\ast }c)$. 
Figs. \ref{F2}a and \ref{F2}b show the magnetic-field dependence
 of the relative MQO amplitudes $\Delta \rho _{xx}/\bar{\rho}
 	_{xx}$ and $\Delta \rho _{xy}/\bar{\rho } _{xy}$ 
 	obtained from the inverse Fourier transformation, where $\bar{\rho} _{xx}$
 	and $\bar{\rho}_{xy}$ are non-oscillating parts of diagonal and Hall magnetoresistivity
 correspondingly.

The Hall-resistivity oscillations are in antiphase
to magnetoresistance oscillations (see Fig. \ref{F2} (a) and (b)), which 
corresponds to the theoretical prediction \cite{Isihara86,GrigorievHallTheory2024} 
in Eqs. (\ref{xx}) and (\ref{xy}). In Fig. \ref{F2}(e) the ratio of the 
absolute values of MQO amplitudes, $\Delta \rho _{xx}/\Delta \rho _{xy}$, 
as a function of $B$ are shown at various temperatures. 
We see that these dependencies are linear at all temperatures in agreement 
with Eq. (\ref{ratio}). The scattering time obtained using Eq. (\ref{ratio}) 
from the slope of this curve at $T=5$ K is $\tau
=(0.99\pm 0.09)10^{-13}$ s, which coincides with the scattering time 
$\tau_\alpha =(0.90\pm 0.07)10^{-13}$ s extracted
from the Dingle plot at the same temperature.

The above results suggest a new and elegant method to determine the electron scattering 
time $\tau $ using the ratio between the MQO amplitudes of diagonal 
and Hall magnetoresistivity. 
To check the applicability region of the proposed method we apply it
at higher temperatures and compare with other common methods. 
As we noted before, the Dingle plots at $T>10$ K demonstrate an unrealistic behavior. 
For $T=5,10,15,20,25$ K these plots are shown in Fig. \ref{F2}(d), 
where all these graphs are almost linear but with a slope that 
continuously decreases with increasing temperature.
This corresponds to the decrease of scattering time $\tau (T)$ with 
increasing temperature, which is unphysical and indicates that 
the L-K formula (\ref{LK}) for the temperature dependence of MQO amplitude 
does not hold. At $T=5$ K the temperature damping factor $R_T$ is only a 
small correction that does not affect the Dingle plot. Hence, the extracted 
Dingle temperature $T_D=13.4$K and the corresponding mean-free time 
$\tau_\alpha \approx 0.9\cdot 10^{-13}$ s are reasonable. 
However, at higher temperature even small violations of the L-K formula 
(\ref{LK}) change dramatically the final Dingle plot and spoil the 
common method of determining $\tau$ from the Dingle plot. 

On contrary, we can extract the scattering time $\tau$ at high temperature from the ratio $\Delta \rho
_{xx}/\Delta \rho _{xy}$. The dependence of this ratio on magnetic field $B$ at $T=5,10,15,20,25,30$ K is
shown in Fig. \ref{F2}(e). In contrast to the Dingle-plot procedure, 
from the ratios $\Delta \rho _{xx}/\Delta \rho _{xy}$ 
we obtain a reasonable temperature dependence of the scattering rate 
$1/\tau (T)$ shown in the inset in Fig. \ref{F2}(e) and given by the sum of contributions from the electron-phonon (e-ph) and
electron-electron (e-e) interaction \cite{Abrik}, 
\begin{equation}
\tau^{-1} (T)=\tau^{-1} (0)+\tau_{e-ph}^{-1} (T)+\tau_{e-e}^{-1} (T),
 \label{tauT}
\end{equation}%
where at low $T<30$K $\tau_{e-ph}^{-1} (T)\propto T^3$ and $ \tau_{e-e}^{-1} (T)\propto T^2$.

With increasing temperature, the MQO in magnetoresistance quickly disappear
according to Eq. (\ref{RT}) due to temperature smearing of the Fermi
level. An increase in temperature also leads to the raise of electron
scattering rate $\tau^{-1}$ because of the e-ph and
e-e interaction \cite{Abrik}. However, in the lowest order of e-ph
interaction and for exponentially weak MQO, the e-ph interaction
leaves the Dingle factor $R_D$ and the effective mass $m^{\ast}$ unchanged 
 in the MQO damping given by Eq. (\ref{LK}) \cite{Fowler65,Engelsberg70}.
This comes from the special cancellation of two terms in the electron self
energy at $T\gg \hbar \omega _{c}$, which enter both $R_D$ and $R_T$. 
Later this cancellation was confirmed
for the 2D electron systems and for the e-e interaction 
\cite{Maslov2003,AGM2006,Chubukov2012} and named the
first Matsubara-frequency rule \cite{Chubukov2012}.

The above cancellation of the $T$-dependence of MQO amplitude \cite{Fowler65,Engelsberg70,Maslov2003,AGM2006,Chubukov2012} 
concerns only the exponential factor given by Eq. (\ref{LK}), which contains the product of $R_T$ and $R_D$.
The prefactors $\omega _{c} \tau $ in Eqs. (\ref{xy}) and (\ref{ratio}), as well as the Dingle factor $R_D$ alone,
do not have this cancellation, and $\tau $ in these prefactors depends on temperature. 

The resistivity $\rho_{xx} (T)$ contains the $T$-dependence of the transport scattering rate $\tau^{-1}_{tr} (T)$, 
which differs from $\tau^{-1} (T)$ at low temperature
\cite{Abrik}. Hence, $\rho_{xx} (T)$ only gives a qualitative dependence $\tau (T)$. 
Thermal conductivity contains $\tau^{-1} (T)$ in combination with the 
electronic part of the specific heat $C(T)\propto T$ \cite{Abrik} 
and also can be used to extract the dependence $\tau^{-1} (T)$.
The temperature dependence of $\tau$ and of the  Dingle factor (\ref{RD})
can also be studied experimentally using the so-called differential 
or slow magnetoresistance oscillations (SlO) 
\cite{KartPRL2002,GrigPRB2003,GrigYBCOPRB2017,Mogilyuk2018,Sinch2016} 
\begin{equation}
\rho _{d}\approx A_{d}\cos \left( 2\pi \Delta F/B\right) R_{D}^{2}
\label{DMO}
\end{equation}%
with a frequency $\Delta F$ proportional not to the Fermi energy 
$\varepsilon _{F}$ or to the Fermi-pocket area but to the splitting 
of electron band structure due to the interlayer transfer integral. 
This energy splitting is not affected by the temperature smearing of the Fermi level, 
hence the SlO do not have the temperature damping factor $R_T$ given by Eq. (\ref{RT}), 
and the temperature damping of SlO is determined only by the electron scattering 
processes entering $\tau^{-1} (T)$. The SlO amplitude is also not affected by the 
macroscopic sample inhomogeneities, which smear the Fermi level and MQO similar to 
temperature \cite{KartPRL2002,GrigPRB2003,GrigYBCOPRB2017,Mogilyuk2018,Grig2012}. 
Therefore, the SlO are often stronger than the usual MQO \cite{KartPRL2002}.
The magnetic intersubband oscillations \cite{Raikh1994,Averkiev2001,TdTIntersubband} 
or difference-frequency oscillations in multiband metals \cite{Leeb2023}
have a similar origin but are less convenient to extract $\tau^{-1} (T)$, since their
amplitudes contain the temperature damping factor $R_T(T)$ 
because the effective masses differ on two different bands or FS pockets. 

We pay attention on the fact which was not seen before: if $\tau $ in
Eq. (\ref{xy}) decreases with temperature, e.g. due to e-e or e-ph
interation, the MQO should fade with temperature much slower for Hall than
for diagonal resistivity, because the oscillatory term in Hall resistivity
is inversely proportional $\tau $. This interesting fact didn`t get enough
attention till now probably because of fact that the work \cite{Isihara86}
was oriented mainly on the quantum Hall effect (QHE) systems. As a rule, QHE
is studied in semiconducting heterostructures having relatively low carrier
concentration. Hence, in these structures the relative MQO in Hall resistance 
appear much weaker than the MQO in magnetoresistance. Another
situation takes place in metallic Q2D compounds where the carrier
concentration is high and the Hall effect is not too large. In such systems one
can expect that the relative MQO in Hall coefficient are much stronger than in diagonal
magnetoresistance. As an indication of such a behavior we notice the first
observation of MQO in high-temperature cuprate superconductors just in the Hall
resistance \cite{ProustNature2007}. From Eq. (\ref{xy}) we see that
the MQO of Hall coefficient are stronger than the MQO of diagonal magnetoresistance 
at low and intermediate magnetic field range when $\omega _{c}\tau \lesssim 1$. 
Thus, for the experimental observation of this enhancement of MQO in Hall 
coefficient in other compounds, the most convenient is to study
Q2D metals in the intermediate magnetic-field range.

To summarize, we performed a comprehensive analysis of the quantum oscillations 
of magnetoresistance tensor in layered rare-earth tritellurides, including the 
intralayer diagonal and Hall magnetoresistance. The magnetic quantum oscillations 
(MQO) of Hall coefficient are much stronger and persist to much higher temperature. 
We show that this is a general effect for MQO in highly anisotropic metals, 
and the combined Hall and diagonal magnetoresistance measurements provide additional 
useful information about the electronic structure. In particular, the ratio of 
MQO amplitudes of diagonal and Hall magnetoresistance components depends linearly 
on the magnetic field, and its slope gives a simpler and much more accurate estimate 
of the electron mean free time than the Dingle plot, especially at finite 
temperature $T\sim \hbar\omega_c$. This provides an elegant new method of measuring 
the electron scattering rate and its temperature dependence in various quasi-2D conductors, 
including high-temperature superconductors, organic metals, layered van-der-Waal crystals, 
topological materials, graphite intercalation compounds, artificial heterostructures, etc.  

The work is supported by RSF-ANR grant RSF-22-42-09018 and ANR-21-CE30-0055.

\end{document}